\documentclass[prd,nofootinbib,preprint,epsfig]{revtex4}
\def\lsim{\mathrel{\rlap{\lower4pt\hbox{\hskip1pt$\sim$}}
    \raise1pt\hbox{$<$}}}      
\def\gsim{\mathrel{\rlap{\lower4pt\hbox{\hskip1pt$\sim$}}
    \raise1pt\hbox{$>$}}}      

\begin{document}

\title{More on deviation from bi-maximal neutrino mixing }

\author{C.~A.~de~S.~Pires \footnote{E-mail: cpires@fisica.ufpb.br}}
\affiliation{Departamento de F\'{\i}sica, Universidade Federal da
Para\'{\i}ba, Caixa Postal 5008, 58051-970, Jo\~ao Pessoa - PB,
Brazil.}


\begin{abstract}
\vspace*{0.5cm}
We study the case of $U^T_l$ presenting the exact bi-maximal mixing form with $U_\nu$ inducing the deviation from the bi-maximal mixing in the final form of the Pontecorvo-Maki-Nakagawa-Sakata neutrino mixing, $U_{PMNS}=U^T_l U_\nu$. We will show that such possibility will lead to a democratic texture for the charged lepton mass matrix and to a neutrino mass matrix with four null entries.
\end{abstract}

\maketitle

\section{Introduction}
\label{sec1}
As a direct consequence of the experimental  observation of neutrino oscillation\cite{oscillation}, neutrinos are massive and the leptonic weak charged current  presents a  mixing matrix\cite{PMNS},
\begin{eqnarray}
{\cal L}_{CC} = \frac{g}{\sqrt{2}}\bar l_L U_{PMNS} \gamma^\mu \nu_L W^-_\mu +H.c.,
\label{CC}	
\end{eqnarray}
where,
 \begin{eqnarray}
U_{PMNS}=U^T_l U_\nu.
	\label{PMNS}
\end{eqnarray}
The matrices that compose $U_{PMNS}$ connect leptons in the flavor basis to leptons in the mass basis
\begin{eqnarray}
	l^{\prime}_L = U_ll_L \,\,\,\,\,\,\,\,\,\, \nu^{\prime}_L = U_\nu \nu_L.
\label{connection}
\end{eqnarray}
In the case of CP-invariance, $U_{PMNS}$  can be parametrized by three angles in the standard way
\begin{eqnarray}
U_{PMNS}=  \left (
\begin{array}{ccccc}
c_{13} c_{12} &\,\,& s_{12} c_{13} &\,\,& s_{13} \\
-s_{12} c_{23}-s_{23} s_{13} c_{12} &\,\,& c_{23} c_{12}-s_{23}
s_{13} s_{12} &\,\,& s_{23} c_{13} \\s_{23} s_{12} -s_{13} c_{23}
c_{12} &\,\,&  -s_{23} c_{12}-s_{13} s_{12} c_{23}&\,\,& c_{23}
c_{13}
\end{array}
\right )\,, \label{CKM} 
\end{eqnarray}
where we have used the short form $c_{ij}\equiv
\cos{\theta_{ij}}$ and $s_{ij}\equiv \sin{\theta_{ij}}$.

Phenomenologically speaking, what matter is  $U_{PMNS}$. However the patterns of $U_l$ and $U_\nu$ are theoretically important because with them we can infer the pattern of the lepton mass matrices  and then understand  the leptonic sector. This is possible because $U_l$  and $U_\nu$  connect the lepton  mass matrices in the mass basis with the flavor basis\footnote{In this work we restrict our discussions to the case of hermitian mass matrices.}
\begin{eqnarray}
	M_l =U_l M^D_l U^T_l , \,\,\,\,\,\,\,\,\,\, M_\nu =U_\nu M^D_\nu U^T_\nu,
	\label{massconnetion}
\end{eqnarray}
where $M^D_l = \mbox{diag}\left ( m_e \,,\, m_\mu \,,\, m_\tau \right)$ and $M^D_\nu = \mbox{diag}\left( m_1 \,,\, m_2 \,,\, m_3 \right)$ are the mass matrices in the mass basis,  while $M_l$  and $M_\nu$  are the mass matrices in the flavor basis. 

Despite of the advance in the experimental neutrino physics, we still do not dispose of  definitive data, which leaves room for much speculation in neutrino physics. It seems that the most definitive result in neutrino physics is that atmospherics neutrino mixing angle is maximal\cite{data}. Regarding solar mixing angle,  we are far from a definitive value, however recent results indicate that, although it is relatively large, it is not maximal\cite{data},
\begin{eqnarray}
0.35 \leq \tan^2(\theta_{sol})\leq 0.52,
\label{solarmixing}	
\end{eqnarray}
while there is only an upper bound on the third angle in (\ref{CKM})\cite{data,chooz}. 

On the other hand, the experiments say nothing about $U_l$  and $U_\nu$, unless we consider neutrinos or charged leptons in a diagonal basis. In this case $U_{PMNS}$ is identified with the lepton in a non-diagonal basis. On the contrary, if we want to know $U_l$  and $U_\nu$ we have to guess their forms. This artifice has been recently employed  to understand the deviation from bi-maximal mixing as required by the recent results in neutrino physics\cite{devbimax}. The idea is to assume neutrino mixing with the bi-maximal form, $U_\nu=U_{\mbox{bimax}}$, and let  $U^T_l$ in charge of the deviation from bi-maximal mixing through the product $U^T_l U_\nu$.

In this work we follow this idea in a different sense. Since what matter is in fact the product $U^T_l U_\nu$, we assume $U^T_l = U_{\mbox{bimax}}$ and let $U_\nu$  in charge of the  deviation from bi-maximal mixing. We also follow a different approach to obtain $U^T_l$ and $U_\nu$. Instead of starting parametrizing  $U_\nu$, we start parametrizing  $U_{PMNS}$. We parametrize $U_{PMNS}$ in a rather synthetic way. With $U_{PMNS}$ at hand we then obtain $U_l^T$  and $U_\nu$ by spliting $U_{PMNS}$ in a product of two matrices. Following this scheme, when we demand $U^T_l =U_{\mbox{bimax}}$, we automatically obtain the form of $U_\nu$. The advantage of such procedure is that it leads to a specific pattern of lepton mass matrices.

Our departing  point is the parametrization of $U_{PMNS}$. When atmospheric mixing angle is taken to be  maximal, the deviation from bi-maximal mixing means the deviation of the solar mixing angle  from maximal by an amount $\delta$. In the case of CP-invariance and $\theta_{13}=0$, this can be described through the following synthetic parametrization of $U_{PMNS}$
\begin{eqnarray}
U_{PMNS}=\left(\begin{array}{ccc} 
 \frac{1}{\sqrt{2}}+\frac{\delta}{\sqrt{2}} & \frac{1}{\sqrt{2}}-\frac{\delta}{\sqrt{2}}  & 0 \\
 -\frac{1}{2}+\frac{\delta}{2} & \frac{1}{2}+\frac{\delta}{2} & 1/\sqrt{2} \\
 \frac{1}{2}-\frac{\delta}{2} & -\frac{1}{2}-\frac{\delta}{2}& 1/\sqrt{2}
\end{array}
\right). 
\label{mixdev}
\end{eqnarray}
In the case of maximal atmospheric mixing angle and $\theta_{13}=0$, the constraint in (\ref{solarmixing}) implies the following range of values to the elements of $U_{PMNS}$
\begin{eqnarray}
U_{PMNS}=\left(\begin{array}{ccc} 
 0.81 - 0.86 & 0.51 - 0.58  & 0 \\
 (-0.36)-(-0.41) & 0.57 - 0.61 & 0.71 \\
 0.36 - 0.41 & (-0.57) - (-0.61)& 0.71
\end{array}
\right). 
\label{mixprediction}
\end{eqnarray}
Confronting (\ref{mixprediction}) with (\ref{mixdev}), we obtain
\begin{eqnarray}
0.14 < \delta < 0.21.
\label{range}
\end{eqnarray}
In summary, the experimental deviation from bi-maximal mixing is compatible with a parametrization by only one parameter.
\section{The case of $U_\nu = U_{bimax}$}

Let us start reviewing the case $U_\nu=U_{\mbox{bimax}}$. First thing to do is to dismember (\ref{mixdev}) in the product, $U^T_l U_\nu$,  such that $U_\nu$ presents the bi-maximal mixing form.  Such dismemberment is possible and unique and leads to
\begin{eqnarray}
	U_{PMNS}= \left(\begin{array}{ccc} 
 1-\delta & -\sqrt{2} \delta  & \sqrt{2} \delta \\
 \frac{\delta}{\sqrt{2}} & 1 & 0 \\
 -\frac{\delta}{\sqrt{2}} & 0 & 1
\end{array}
\right)\left(\begin{array}{ccc} 
 \frac{1}{\sqrt{2}} & \frac{1}{\sqrt{2}} & 0 \\
 -\frac{1}{2} & \frac{1}{2}  & \frac{1}{\sqrt{2}} \\
 \frac{1}{2}  & -\frac{1}{2} & \frac{1}{\sqrt{2}}
\end{array}
\right). 
\label{spliting}
\end{eqnarray}
By comparing (\ref{PMNS}) and (\ref{spliting}) we recognize
\begin{eqnarray}
	U^T_l=\left(\begin{array}{ccc} 
 1-\delta & -\sqrt{2} \delta  & \sqrt{2} \delta \\
 \frac{\delta}{\sqrt{2}} & 1 & 0 \\
 -\frac{\delta}{\sqrt{2}} & 0 & 1
\end{array}
\right),\,\,\,  \mbox{and} \,\,\, U_\nu = \left(\begin{array}{ccc} 
 \frac{1}{\sqrt{2}} & \frac{1}{\sqrt{2}} & 0 \\
 -\frac{1}{2} & \frac{1}{2}  & \frac{1}{\sqrt{2}} \\
 \frac{1}{2}  & -\frac{1}{2} & \frac{1}{\sqrt{2}}
\end{array}
\right). 
\label{recog}
\end{eqnarray}

As we discussed above, this possibility is interesting because $U_\nu =U_{\mbox{bimax}}$ implies in a very well known texture of the neutrino mass matrix
\begin{eqnarray}
M_\nu = \left(\begin{array}{ccc} 
 \frac{1}{2}(m_1+m_2) & \frac{1}{2\sqrt{2}}(m_2-m_1) &  \frac{1}{2\sqrt{2}}(m_1-m_2) \\
 \frac{1}{2\sqrt{2}}(m_2-m_1) &  \frac{1}{4}(m_1+m_2)+\frac{m_3}{2} & -\frac{1}{4}(m_1+m_2)+\frac{m_3}{2} \\
 \frac{1}{2\sqrt{2}}(m_1-m_2) & -\frac{1}{4}(m_1+m_2)+\frac{m_3}{2} &  \frac{1}{4}(m_1+m_2)+\frac{m_3}{2}
\end{array}
\right). 
\label{bimaxtexture}	
\end{eqnarray}
The interesting point behind such texture is that in the case of inverted hierarchy  it manifests the non-standard $L_e -L_\mu -L_\tau$ symmetry\cite{symmetry}. 

On the charged lepton sector, the splitting above leads to the following texture for the charged lepton mass matrix
\begin{eqnarray}
M_l \approx \left(\begin{array}{ccc} 
 -\frac{m_\tau}{2} \delta^2 & \frac{m_\mu}{\sqrt{2}}\delta &  \frac{m_\tau}{\sqrt{2}}\delta \\
 \frac{m_\mu}{\sqrt{2}}\delta&  m_\mu & 0 \\
 \frac{m_\tau}{\sqrt{2}}\delta & 0 &  m_\tau
\end{array}
\right). 
\label{CLM1}
\end{eqnarray}
With the range of values for $\delta$ given above, we see that this matrix presents a mild hierarchical structure. 
 
\section{The case of $U^T_l =U_{bimax}$}

Another interesting possibility arises when we  attribute the bi-maximal mixing form to $U^T_l$ and let $U_\nu$ in charge of generating the deviation from bi-maximal mixing.  In this case the unique way of dismembering (\ref{mixdev}) is this
\begin{eqnarray}
	U_{PMNS}= \left(\begin{array}{ccc} 
 \frac{1}{\sqrt{2}} & \frac{1}{\sqrt{2}} & 0 \\
 -\frac{1}{2} & \frac{1}{2}  & \frac{1}{\sqrt{2}} \\
 \frac{1}{2}  & -\frac{1}{2} & \frac{1}{\sqrt{2}}
\end{array}
\right)\left(\begin{array}{ccc} 
 1 & -\delta  & 0 \\
 \delta & 1 & 0 \\
 0 & 0 & 1
\end{array}
\right). 
\label{invertspliting}
\end{eqnarray}

Confronting this product with (\ref{PMNS}), we recognize
\begin{eqnarray}
	U^T_l=\left(\begin{array}{ccc} 
 \frac{1}{\sqrt{2}} & \frac{1}{\sqrt{2}} & 0 \\
 -\frac{1}{2} & \frac{1}{2}  & \frac{1}{\sqrt{2}} \\
 \frac{1}{2}  & -\frac{1}{2} & \frac{1}{\sqrt{2}}
\end{array}
\right),\,\,\,\ \mbox{and}\,\,\,\,\,\,\,  U_\nu = \left(\begin{array}{ccc} 
 1 & -\delta  & 0 \\
 \delta & 1 & 0 \\
 0 & 0 & 1
\end{array}
\right). 
\label{recoginvert}
\end{eqnarray}
The interesting point behind such splitting of $U_{PMNS}$ is the textures for the lepton mass matrices that $U^T_l$  and $U_\nu$  imply. Considering first charged leptons
\begin{eqnarray}
M_l= \left(\begin{array}{ccc} 
 \frac{m_e}{2}+\frac{m_\mu}{4}+ \frac{m_\tau}{4} & \frac{m_e}{2}-\frac{m_\mu}{4}- \frac{m_\tau}{4} &  -\frac{m_\mu}{2\sqrt{2}}+\frac{m_\tau}{2\sqrt{2}} \\
 \frac{m_e}{2}-\frac{m_\mu}{4}- \frac{m_\tau}{4}&  \frac{m_e}{2}+\frac{m_\mu}{4}+ \frac{m_\tau}{4} & \frac{m_\mu}{2\sqrt{2}}-\frac{m_\tau}{2\sqrt{2}}  \\
 -\frac{m_\mu}{2\sqrt{2}}+\frac{m_\tau}{2\sqrt{2}}  & \frac{m_\mu}{2\sqrt{2}}-\frac{m_\tau}{2\sqrt{2}}  &  \frac{m_\mu}{2}+\frac{m_\tau}{2} 
\end{array}
\right).
\label{newchargedleptontexture}
\end{eqnarray}
Perceive that $M_l$ presents a democratic texture where all the  entries have the same order of magnitude dictated by the presence of $m_\tau$ in every matrix element. To see this better let us substitute the charged lepton masses that appear there by their respective values( $m_e = 0.51 $ MeV , $m_\mu = 107$ MeV  and $m_\tau = 1.77$ GeV). With this we obtain
\begin{eqnarray}
&&M_l=\left(\begin{array}{ccc} 
 0.47 & -0.47 &  0.59 \\
 -0.47& 0.47 & -0.59  \\
 0.59  & -0.59  &  0.94
\end{array}
\right)\mbox{GeV}.
\label{prediction}
\end{eqnarray}
The interesting point behind such democratic texture is that it implies uniform Yukawa couplings regarding their  order of magnitude. For example, for a VEV of order of $10^2$ GeV, we have Yukawa couplings of order of $10^{-3}$. This makes such scenario phenomenologically appealing since there is no significant hierarchy between the Yukawa couplings.

Considering neutrinos, the form of $U_\nu$ in (\ref{recoginvert}) leads to a very simple texture for the neutrino mass matrix\cite{WB}
\begin{eqnarray}
M_\nu= \left(\begin{array}{ccc} 
m_1  &  (m_1-m_2) \delta & 0 \\
(m_1-m_2)\delta & m_2  & 0 \\
0 & 0 & m_3
\end{array}
\right).
\label{newneutrinotexture}
\end{eqnarray}
Note that   its 11-entry  is not zero, which support neutrinoless 2$\beta$ decay, and that it allows normal hierarchy as well as inverted hierarchy.  

Let us discuss a little how to originate the textures in (\ref{newchargedleptontexture})  and (\ref{newneutrinotexture}). The central point is to generate the four texture zeros in the neutrino mass matrix. Perhaps the faster, but not the most economic, manner of generating texture zeros is through the method developded in Ref. \cite{lavoura}. According to that method the zero entries are enforced by means of Abelian symmetries. It is assumed that each of lepton multiplets $f$ transforms under a separate Abelian group ${\cal G}(f)$  with $f=l_{aR} , L_{aL}\,\,\, (a=1,2,3)$. Then for each non-zero entry in $M_l$  and $M_\nu$ we need to introduce a Higgs multiplets with appropriate transformation properties under ${\cal G}=\times_f{\cal G}(f)$. The texture zeros  is obtained simply by not introducing into the theory the corresponding scalar multiplet. We follow such method in conjunction with the type II see-saw mechanism in order to get the textures in (\ref{newchargedleptontexture})  and (\ref{newneutrinotexture}). For this we have to introduce six Higgs doublets, $\phi_{ab} $, transforming as $\phi_{ab} : {\cal G}^*(l_{aR}) \otimes {\cal G}(L_{bL})$ , and four Higgs triplets, $\Delta_{ij}$, transforming as $\Delta_{ij} : {\cal G}^*(L_{iL}) \otimes {\cal G}^*(L_{jL})$. The dangerous Goldstones of such symmetries are avoided by allowing terms  in the scalar  potential that break them softly.

 The  invariant  Yukawa interactions that we can form with such set of particle content is
\begin{eqnarray}
	{\cal L}= \sum_{a,b} h_{ab}\bar L_{aL} l_{aR}\phi_{ab} + \sum_i g_{ii}\bar{L^c_{iL}} \Delta_{ii} L_{iL} + g_{12}\bar{L^c_{1L}} \Delta_{12} L_{2 L}+H.c.
	\label{lagrangean}
\end{eqnarray}
 When the Higgs multiplets $\phi_{ab}$  and $\Delta_{ij}$  develop their respective VEVs $v_{\phi_{ab}}$  and $v_{\Delta_{ij}}$, the Lagrangean above leads to the required textures in (\ref{newchargedleptontexture})  and (\ref{newneutrinotexture}). 
 
 The values of the entries for $M_l$  in (\ref{prediction})  is obtained by a fair manipulation of the Yukawa couplings $h_{ab}$  and the VEVs $v_{\phi_{ab}}$. This method allows the interesting possibility of having  $h_{ab}=1$ with $v_{\phi_{ab}}$ being responsible by the respective entries in (\ref{prediction}).
 
In regard to neutrino masses, their smallness requires tiny value for each $v_{\Delta_{ij}}$. This is obtained through the type II see-saw mechanism. This is a generic mechanism, which can be
illustrated by the following simple model that has only one $\phi$ and one
$\Delta$. Consider the following Higgs potential for this
system~\cite{moh1}:
\begin{eqnarray}
V(\phi, \Delta) &=& M^2 \Delta^{\dagger}\Delta -\mu^2\phi^{\dagger} \phi 
+ \lambda_{\phi} (\phi^{\dagger}\phi)^2   
+ \lambda_{\Delta}\Delta^{\dagger}\Delta  \nonumber \\
&&+\lambda_{\phi\Delta}\Delta^{\dagger}\Delta\phi^{\dagger}\phi +
M_{\Delta\phi\phi}\Delta^{\dagger}\phi\phi ~ + ~ H.c.
\label{seesaw}
\end{eqnarray}
Let us choose $\mu\sim 10^2$ GeV and $M\sim M_{\Delta\phi\phi}\gg \mu$; in
this case, the VEV of $v_{\phi}\approx \mu$ whereas the VEV of 
$v_{\Delta} \sim \frac{\mu^2}{M}\ll \mu$.
This mechanism has been labeled type II see-saw and we see that
if $M\simeq 10^{14}$ GeV, then we get $v_{\Delta}\simeq 0.1$ eV. 
In the presence of more $\Delta$ fields and extra symmetries that our model 
has this mechanism still operates\cite{morefields}.

In summary, in this short note we parametrized the deviation from bi-maximal mixing form of $U_{PMNS}$ with only one parameter. Next we dismembered such matrix in the product of other two, $U^T_l$ and $U_\nu$. We then reviewed  the case when $U_\nu$  presents the exact bi-maximal form with $U^T_l$ being responsible by the deviation from the exact bi-maximal mixing. Next we inverted the role played by $U^T_l$  and $U_\nu$ where now $U^T_l$ has the exact bi-maximal mixing form, while $U_\nu$ is now responsible by the deviation from bi-maximal mixing. This leads to a democratic texture for the  charged leptons and to a  simple texture for the neutrino mass matrix with four null entries. We finish by stating that taking charged leptons in a non-diagonal basis is not only a plausible possibility but may reveal itself as a clean and interesting option in what concern neutrino physics.

\acknowledgments
This work was supported by Conselho Nacional de Desenvolvimento
Cient\'{\i}fico e Tecnol\'ogico (CNPq).


\begin{references}

\bibitem{oscillation}
SNO collaboration, Q.R. Ahmad {\it et al.},
Phys. Rev. Lett. {\bf 89}, 011301 (2002); Phys. Rev. Lett. {\bf 89}, 011302 (2002);
 Super-Kamiokande Collaboration (Y.~Fukuda {\it et
al.}), Phys.~Lett. B{\bf 436}, 33 (1998); Phys.~Rev.~Lett. {\bf
81}, 1158 (1998); Erratum {\bf 81}, 4279 (1998); Phys.~Rev.~Lett.
{\bf 81}, 1562 (1998); Phys.~Rev.~Lett. {\bf 82}, 1810 (1999);
Super-Kamiokande collaboration, Y. Suzuki, Nucl.~Phys. B{\bf 91}
(Proc.~Suppl.), 29 (2001); Super-Kamiokande collaboration, S.
Fukuda et al., Phys.~Rev.~Lett. {\bf 86}, 5651 (2001);
KamLAND Collaboration, K. Eguchi {\it et al.}, Phys. 
Rev. Lett. {\bf 90}, 021802 (2003).


\bibitem{PMNS}
B. Pontecorvo,  Sov. Phys. JETP {\bf 6}, 429 (1957), Zh. Eksp. Teor. Fiz. {\bf 33}, 549 (1957);
B. Pontecorvo,  Sov.Phys.JETP {\bf 7}, 172 (1958), Zh. Eksp. Teor. Fiz. {\bf 34},  247 (1957) ;
 Z. Maki, M. Nakagawa, S. Sakata,  Prog. Theor. Phys. {\bf 28}, 870 (1962). 
 
 
 \bibitem{data}
  A. Bandyopadhyay, S. Choubey, S. Goswami, S. T. Petcov, D. P. Roy,  Phys. Lett. B{\bf 583}, 134 (2004);
  G.~L.~Fogli, E.~Lisi, A.~Marrone, D.~Montanino, A. Palazzo, and  A.~M.~Rotunno, Phys. Rev. D{\bf 67}, 073002 (2003); 
G.~L. Fogli, E.~Lisi, A.~Marrone, and D.~Montanino,   Phys. Rev. D{\bf 67}, 093006 (2003);
H. Nunokawa, W. J. C. Teves, R. Zukanovich Funchal,
Phys. Lett. B {\bf 562}, 28 (2003); 
P. C. de Holanda, A. Y. Smirnov, JCAP {\bf 0302}, 001 (2003).

\bibitem{chooz}
CHOOZ Collaboration, M. Apollonio {\it et al.}, Phys. Lett. B {\bf 420}, 
397 (1998);
 F. Boehm {\it et al}, Phys. Rev. Lett. {\bf84},  3764 (2000). 



\bibitem{devbimax}
 M. Raidal, hep-ph/0404046;
 W. Rodejohann, hep-ph/0403236;
 A. Romanino, hep-ph/0402258;
 G. Altarelli, F. Feruglio, I. Masina, hep-ph/0402155;
 A. de Gouvea, hep-ph/0401220 ;
 P. H. Frampton, S. T. Petcov, W. Rodejohann, hep-ph/0401206 ;
 W. Rodejohann, Phys. Rev. D {\bf 69}, 033005 (2004); 
 C. Giunti, M. Tanimoto,  Phys. Rev. D{\bf 66}, 11300 (2002);
 C. Giunti, M. Tanimoto, Phys. Rev. D{\bf 66},  053013 (2002);
 T. Ohlsson, G. Seidl, Nucl. Phys. B{\bf 643}, 247 (2002);
 Z. z. Xing,  Phys. Rev. D{\bf 64}, 093013 (2001). 

 
 \bibitem{symmetry}
 S. T. Petcov, Phys. Lett. B{\bf 110}, 245 (1982). 
R. Barbieri, L. Hall, A. Strumia and N. Weiner, JHEP 9812 (1998) 017;
R. Barbieri, L. J. Hall, A. Strumia, Phys.Lett. {\bf B445}, 407 (1999);
Y. Grossman, Y. Nir and Y. Shadmi, JHEP 9810, 007 (1998);
M. Jezabek, Y. Sumino,  Phys. Lett. {\bf B457}, 139 (1999);
A. Joshipura and S. Rindani, Phys.Lett. B464, 239 (1999); Eur. Phys. J. C{\bf 14}, 85 (2000); R. N. Mohapatra, A. Perez-Lorenzana and C. A. de S. Pires, 
Phys. Lett. B{\bf 474}, 355 (2000);
 H. S. Goh, R. N. Mohapatra, S. P. Ng, Phys. Lett. B {\bf 542}, 116 (2002);
  K. S. Babu, R. N. Mohapatra, Phys. Lett. B{\bf 532}, 77 (2002);
 G. Altarelli, F. Feruglio,  hep-ph/0206077.
 
 \bibitem{WB}
 It is important to stress that the textures for $M_l$  and $M_\nu$ given in Eqs. (\ref{newchargedleptontexture})  and (\ref{newneutrinotexture}) are fundamental in the sense that they are not obtained from a more general set of mass matrices by a weak basis transformation. For more details about weak basis transformation see: G. C. Branco, D. Emmanuel-Costa and R. Gonzalez Felipe, Phys. Lett. B{\bf 477}, 147 (2000).
 
 \bibitem{lavoura}
  W. Grimus, A. S. Joshipura , L. Lavoura, M. Tanimoto, hep-ph/0405016.

 
\bibitem{moh1} 
R. N. Mohapatra and G. Senjanovi\'c, Phys. Rev. {\bf D 23}, 165 (1981); 
C. Wetterich, Nuc. Phys. {\bf B 187}, 343 (1981);
E. Ma and Sarkar, Phys. Rev. Lett. {\bf 80 }, 5716  (1998).

\bibitem{morefields}
R. N. Mohapatra, A. P. Lorenzana, C. A. de S. Pires, Phys. Lett. B{\bf 474}, 355 (200).
 
 
  

 








\end{references}
\end{document}